# Temporal-spatial Adaptation of Promptable SAM Enhance Accuracy and Generalizability of cine CMR Segmentation

Zhennong Chen[*], Sekeun Kim[*], Hui Ren, Quanzheng Li, Xiang Li

*Center of Advanced Medical Computing and Analysis, Massachusetts General Hospital and Harvard Medical School, Boston, MA, 02114, USA*
`xli60@mgh.harvard.edu`
\* *co-first authors*

**Abstract.** Accurate myocardium segmentation across all phases in one cardiac cycle in cine cardiac magnetic resonance (CMR) scans is crucial for comprehensively cardiac function analysis. Despite advancements in deep learning (DL) for automatic cine CMR segmentation, generalizability on unseen data remains a significant challenge. Recently, the segment-anything-model (SAM) has been invented as a segmentation foundation model, known for its accurate segmentation and more importantly, zero-shot generalization. SAM was trained on two-dimensional (2D) natural images; to adapt it for comprehensive cine CMR segmentation, we propose cineCMR-SAM which incorporates both temporal and spatial information through a modified model architecture. Compared to other state-of-the-art (SOTA) methods, our model achieved superior data-specific model segmentation accuracy on the STACOM2011 when fine-tuned on this dataset and demonstrated superior zero-shot generalization on two other large public datasets (ACDC and M&Ms) unseen during fine-tuning. Additionally, we introduced a text prompt feature in cineCMR-SAM to specify the view type of input slices (short-axis or long-axis), enhancing performance across all view types.

**Keywords:** Cardiac magnetic resonance, Segmentation, Foundational Model

## 1 Introduction

Accurate and reproducible assessment of the myocardium is crucial to indicate previous infarcts, cardiomyopathies, or inflammatory diseases[1]. Cine cardiac magnetic resonance (CMR) imaging is considered as the gold standard modality for myocardial anatomy and function, but to quantify these metrics requires manual segmentation, a labor-intensive task hampered by variations in image quality and observer education and experience[2]. Moreover, conditions like cardiac dyssynchrony and diastolic heart failure demand segmentation across all cardiac phases for a more thorough temporal analysis, significantly extending segmentation time. Thus, there is a critical need for an automated cine CMR segmentation method capable of accurately segmenting all cardiac cycle phases in clinical practice.

There have been advancements in DL for automatic cine CMR segmentation. Specifically, cine CMR segmentation here refers to "2D+T" CMR segmentation, which involves segmenting myocardium in one slice across all cardiac phases *simultaneously*. This requires DL models to understand both spatial information and temporal dynamics and maintain temporal consistency across all cardiac phases. Current mainstream approaches incorporate temporal information in cine CMR using three ways: (1) adding recurrent layers in a 2D convolutional neural network (CNN) to encode sequential information along the temporal axis[3], [4], [5], (2) utilizing a 3D CNN with 3D kernels that treat the temporal axis as a depth dimension[6], and (3) employing attention mechanisms on both temporal and spatial axes[7], [8], [9]. There are also studies[9], [10] that have addressed volumetric 3D+T segmentation, but we focus on 2D+T segmentation since segmenting in 2D slices allows networks to work with images even if they have different slice thicknesses or severe inter-slice misalignment due to cardiac or respiratory motion[11]. Despite these advancements, a major challenge remains in generalizability to unseen datasets. Several research[12], [13], [14] have demonstrated significant performance degradation when models are applied to new, unseen datasets from different centers or vendors. Another limitation is most DL methods are only validated on short-axis (SAX) views[15], but segmentation in long-axis (LAX) views is critical as it provides clinically valuable parameters such as longitudinal strains.

Recently, Segment-Anything-Model (SAM)[16] has been introduced as a segmentation foundation model trained on one-billion-image dataset. It is renowned for its segmentation accuracy and more importantly, zero-shot generalization and user-defined prompt input. Fine-tuning SAM on general medical images has outperformed specialist CNN[17], [18], and adapting SAM to specific image modalities has further enhanced accuracy considerably[19], [20]. However, there is no study tailoring SAM for cine CMR segmentation to leverage its generalization ability and prompt inputs. Therefore, we propose cineCMR-SAM with specific model modifications for cine CMR.

Our contributions are three-folds. First, we integrate the 2D SAM model for 2D+T CMR segmentation. Specially, we incorporate time-space self-attention in the SAM vision transformer (ViT) encoders, enabling extraction of both temporal and spatial information. Second, we design a text prompt feature to specify the view type of input slices (SAX or LAX) to enhance the segmentation across all views. Third, we fine-tune our model on one multi-center, multi-vendor public dataset (STACOM2011) and then demonstrate its superior zero-shot generalization compared to state-of-the-art (SOTA) methods on two other large datasets (ACDC and M&Ms) unseen during fine-tuning.

## 2   Methods

Fig. 1 illustrates our model. In this section, we first present how we integrate a 2D SAM model for 2D+T segmentation. Then we introduce the prompt feature to specify the view type. Lastly, we cover some other model modifications and training procedures.

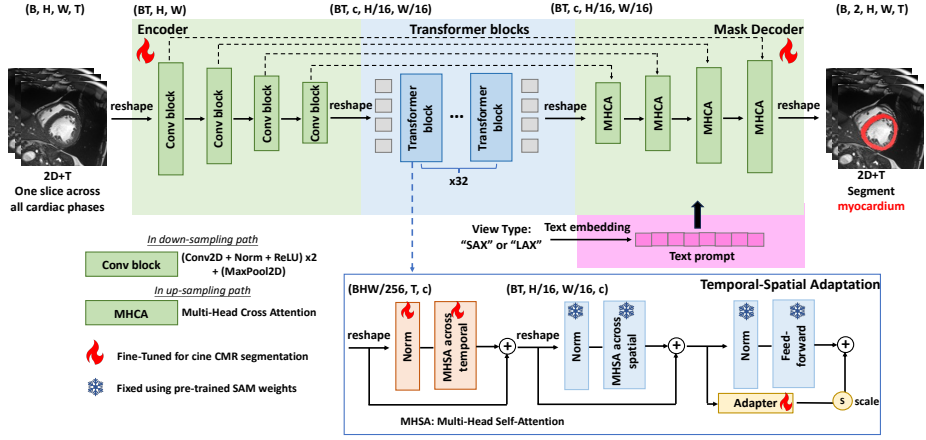

**Fig. 1.** The overall model structure. It includes (1) refined SAM ViT blocks with temporal-spatial adaptation (blue section); (2) U-Net framework as multi-scale encoder and decoder (green section); (3) text prompts to specify the view type (pink section). Components undergoing fine-tuning for CMR segmentation are denoted by a "fire" icon, whereas those retaining pre-trained SAM weights are marked with a "snowflake" icon. The data dimension before input into each module is annotated at the top of the respective module, where B = batch size, c = feature channel number, H and W = height and width of the input slice, and T = the number of cardiac phases.

### 2.1 Integrate 2D SAM for 2D+T Segmentation

The SAM initially has three main parts: an image encoder, a prompt encoder and a mask decoder, all pre-trained on a vast one-billion-image dataset. The image encoder features a series of ViT blocks, critical for SAM's effectiveness in downstream tasks. Our cineCMR-SAM uses the "vit-h' variant with 32 transformer blocks. The initial design of SAM ViT blocks is for 2D images, extracting only spatial information using a multi-head self-attention (MHSA) mechanism. To adapt it for 2D+T segmentation, we were inspired by TimeSformer (the space-time self-attention module)[7] and thus refine the ViT blocks to have temporal MHSA and spatial MHSA sequentially applied one after the other. Skip connections are used to integrate temporal-spatial information. The layers in one refined ViT block are shown in the blue box in Fig. 1.

Regarding data dimension change through the model framework, for model inputs, we extract a CMR slice spanning across one cardiac cycle $x = \{x_t\}_{t=1}^{T}$, $x \in \mathbb{R}^{B \times H \times W \times T}$. Here, $B$ denotes the batch size, $H \times W$ denote the slice dimensions and $T$ denotes the number of phases in one cardiac cycle. Before the inputs are passed into the SAM backbone, a reshape operation is applied to transform $x \in \mathbb{R}^{B \times H \times W \times T}$ into $x \in \mathbb{R}^{BT \times H \times W}$ by merging the phases into the batch dimension. Then for ViT feature extraction, prior to feeding into the temporal MHSA, they are reshaped from $[BT, H/16, W/16, c]$ to $[BWH/256, T, c]$ so that the attention works on the temporal axis. Here $c$ denotes feature channel number, while $H/16$ and $W/16$ denote the spatial dimensions of feature maps, which are down-sampled by 16 times because of the patch embedding process in transformer. After temporal MHSA, we transform the dimension from [*BWH*/256, *T*, *c*] back to [*BT*, *H*/16, *W*/16, *c*] for spatial MHSA. Using these dimension transformation operations, we can apply attention on both the temporal and spatial axes.

### 2.2 Text Prompt to Specify View Type

SAM allows users to input prompts to guide segmentation. Our goal is to enable cineCMR-SAM to segment both SAX and LAX using a *single* model. We hypothesize that leveraging the text prompt feature to specify the view type of the model input can enhance segmentation accuracy across all views. Concretely, we use text "SAX" and "LAX" as text prompts for SAX and LAX view inputs respectively. The text is embedded using the trainable prompt encoder and integrated into the decoder part of the model. Currently the prompt context is defined manually for each input, but it can be automated using deep learning[21].

### 2.3 Other Modifications and Training Procedure

As shown in Fig. 1, we utilize a 2D U-Net framework for our model and incorporate the refined SAM's ViT blocks at the bottom of U-Net. We hypothesize that using U-Net can better extract multi-scale spatial features in the image while the transformer can learn the long-range dependencies among pixels[22], [23]. Inspired by U-Net Transformer[8], each stage of the U-Net's upsampling pathway is enhanced by incorporating a multi-head cross-attention (MHCA) module, which reduces noise and irrelevant elements in the skip-connected features.

For parameter-efficient transfer learning, we apply an adapter module after the feature extraction in each ViT block and scale its output using a scaling factor (empirically $s = 0.5$) to balance the task-agnostic features and the task-specific features[24]. During model fine-tuning, we freeze the pre-trained SAM weights for spatial MHSA and feed-forward layers (marked by the "snowflake" icon in Fig. 1) to leverage SAM's inherent generalizability abilities, while fine-tuning all other parts of the model including the temporal MHSA, adapter module and U-Net layers. Our model is trained to segment myocardium in a CMR slice spanning across one cardiac cycle *simultaneously*. For image preprocessing 2D+T data, the input slice is center cropped to [H, W] = [128, 128]. Image intensities were normalized to [0,1] via min-max normalization method. The model was trained and tested on a single DGX-A100 GPU (NVIDIA, CA, USA) and consumed around 12GB of GPU memory.

## 3 Experiments

We used a multi-center, multi-vendor cine CMR dataset (STACOM2011) to evaluate the segmentation accuracy of our data-specific model. To validate the zero-shot generalizability, we fine-tuned our model on the STACOM2011 and then applied it zero-shot on two other large datasets (ACDC and M&Ms) that were unseen during fine-tuning.

**Fine-tuning Dataset.** We utilized the STACOM 2011 dataset[25], which is a multi-center, multi-vendor dataset comprising 100 patients with coronary artery disease and prior myocardial infarction. MR vendors used include GE, Philips and Siemens. Pixel-wise segmentation of the myocardium for all cardiac phases is publicly available as ground truth. Both SAX and LAX images are included. We selected this dataset since it is the only public dataset with ground truth labels available for all cardiac phases, making it suitable for training a 2D+T segmenter. In the experiments, we first split the dataset into 60:40 for training/validation and testing to evaluate segmentation accuracy of the data-specific models. Then we fine-tuned the model on the entire STACOM for zero-shot generalization evaluation on two testing datasets.

**Testing Dataset.** We utilized the ACDC dataset[26] and the M&Ms dataset[14] for zero-shot generalization evaluation. The ACDC dataset is a single-center, single-vendor dataset composed of 100 cine CMR cases from five different pathological groups. All images were from a single center using Siemens scanners. The M&Ms dataset is a multi-center, multi-vendor dataset composed of 136 cases in its testing cohort from 5 medical centers (4 from Spain, 1 from Germany) and 4 MR vendors (GE, Philips, Siemens and Canon). This cohort has a wide collection of cardiovascular diseases. In both datasets, the ground truth labels are available for only end-diastole (ED) and end-systole (ES) phases in SAX slices.

**Comparison with SOTA methods.** We compared our method with three representative SOTA methods to address 2D+T cine CMR segmentation. DeepIED[4] (noted as "2D recurrent" in Table 1 and 2) represents adding recurrent layers in a 2D CNN to encode sequential information along the temporal axis. nnUNet-3D[27] represents utilizing a 3D CNN with 3D kernels that treat the temporal axis as a depth dimension. U-transformer[8], which shares the same U-Net framework design as ours, represents employing attention mechanisms. We added the temporal attention to the original U-transformer to enable temporal-spatial information extraction. Note we only fine-tuned and evaluated the models on SAX images and our model was free of prompts in these comparison studies, leaving the LAX and prompts to the next section.

We first trained all methods on STACOM2011 using 60:40 split (60 training/validation, 40 testing) to evaluate the data-specific model performance. Table 1 shows the Dice coefficient and Hausdorff distance (HD). We show that our cineCMR-SAM significantly outperforms ($p<0.001$ by one-tailed Wilcoxon signed-rank test) the CNN-based methods (2D recurrent and nnUNet3D) as well as the fusion of U-Net and transformer (U-transformer) for all levels of SAX slices (basal, mid and apical). We then fine-tuned all methods using the entire STACOM2011 dataset and evaluated zero-shot generalization on ACDC and M&Ms datasets. Table 2 shows that our method has significantly better generalization ability in both ACDC (Dice = 0.850 and HD = 4.066 pixels for myocardium in all SAX slices, $p<0.001$ by one-tailed Wilcoxon signed-rank test) and M&Ms (Dice = 0.835 and HD = 4.674 pixels for myocardium in all SAX slices, $p<0.001$) datasets compared to all other methods. Especially, Fig. 2 shows that in M&Ms dataset there were no significant differences in our models's performance across all vendors and centers using one-way analysis of variance (ANOVA), except for vendor 1 used in center 1 with lower performance ($p<0.001$ by Tukey's range test).

**Table 1.** Results of data-specific models on STACOM Dataset. The dataset was split 60:40 and the latter were for testing. HD = Hausdorff Distance (unit: pixel)

| Methods | All SAX | | Basal SAX | | Mid SAX | | Apical SAX | |
|---|---|---|---|---|---|---|---|---|
| Metrics | Dice | HD | Dice | HD | Dice | HD | Dice | HD |
| 2D Recurrent[4] | 0.865 | 3.842 | 0.877 | 3.214 | 0.874 | 3.126 | 0.813 | 3.811 |
| nnUnet3D[27] | 0.877 | 3.867 | 0.888 | 3.221 | 0.882 | 3.345 | 0.839 | 3.293 |
| U-transformer[8] | 0.880 | 3.622 | 0.890 | 3.044 | 0.887 | 2.874 | 0.836 | 3.486 |
| Ours | **0.890** | **2.853** | **0.897** | **2.471** | **0.898** | **2.411** | **0.851** | **2.824** |

**Table 2.** Results of zero-generalization on ACDC and M&Ms dataset. The models were fine-tuned on the entire STACOM dataset and directly applied to new datasets.

| Methods | All SAX | | Basal SAX | | Mid SAX | | Apical SAX | |
|---|---|---|---|---|---|---|---|---|
| *ACDC* | Dice | HD | Dice | HD | Dice | HD | Dice | HD |
| 2D Recurrent[4] | 0.809 | 6.720 | 0.825 | 6.344 | 0.811 | 6.181 | 0.757 | 5.917 |
| nnUnet3D[27] | 0.785 | 7.416 | 0.806 | 6.482 | 0.785 | 6.257 | 0.729 | 6.880 |
| U-transformer[8] | 0.828 | 4.741 | 0.846 | 4.321 | 0.829 | 4.092 | 0.790 | 4.264 |
| Ours | **0.850** | **4.066** | **0.871** | **3.202** | **0.844** | **3.776** | **0.793** | **3.788** |
| *M&Ms* | | | | | | | | |
| 2D Recurrent[4] | 0.786 | 6.525 | 0.793 | 7.530 | 0.786 | 7.079 | 0.763 | 6.184 |
| nnUnet3D[27] | 0.801 | 6.231 | 0.806 | 5.028 | 0.810 | 5.405 | 0.750 | 6.081 |
| U-transformer[8] | 0.810 | 5.611 | 0.803 | 6.397 | 0.820 | 5.146 | 0.781 | 4.732 |
| Ours | **0.835** | **4.674** | **0.839** | **4.045** | **0.844** | **3.974** | **0.797** | **4.498** |

**Table 3.** Effectiveness of prompt features. We fine-tuned both models (cineCMR-SAM without or with text prompts) on the first 60% of STACOM and tested on the rest 40%.

| Prompts | All SAX | | All LAX | |
|---|---|---|---|---|
| | Dice | HD | Dice | HD |
| without text | 0.881 | 3.133 | 0.837 | 6.264 |
| with text | **0.890** | **2.909** | **0.845** | **5.440** |

Notably, vendor 4 (Canon) belongs to a vendor that was not included in the fine-tuning dataset, yet our model achieved performance comparable to that with vendors seen in the fine-tuning dataset.

**Effectiveness of text prompts.** In this study, we enabled the text prompt feature and assessed its effectiveness to improve segmentation across all types of CMR views. Concretely, we evaluated two versions of cineCMR-SAM: one with text prompts and the other without. We split STACOM2011 into a 60:40 ratio and fine-tuned the models on both SAX and LAX images together. Table 3 shows that enabling text prompts in our model increases the Dice coefficient from 0.881 to 0.890 in SAX and from 0.837 to 0.845 in LAX, while decreases the HD from 3.133 pixels to 2.909 in SAX and from 6.264 to 5.440 in LAX. These results indicate that specifying the view type of input slices enhances segmentation across all views. Notably, compared to the values reported in the section above (i.e., cineCMR-SAM without prompts trained only on SAX images), cineCMR-SAM without prompts trained on both view types shows a slight drop in performance in SAX (Dice drops 0.890 to 0.881), likely due to the model needing to learn from different input types. After using prompts, the performance returns to the same level (p=0.167 by two-tailed Wilcoxon signed rank test).

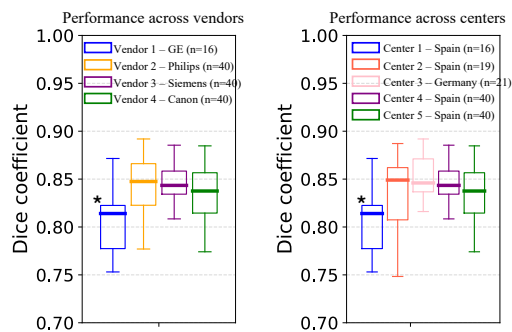

**Fig. 2.** The zero-shot generalization of our model across different vendors and centers in M&Ms dataset. The data are from 4 vendors and 5 centers, where center 1 uses vendor 1, center 2 and 3 uses vendor 2, center 4 uses vendor 3 and center 5 uses vendor 4. The "*" represent that Dice in vendor 1/center1 is statistically lower compared to the others ($p<0.001$ by Tukey's range test).

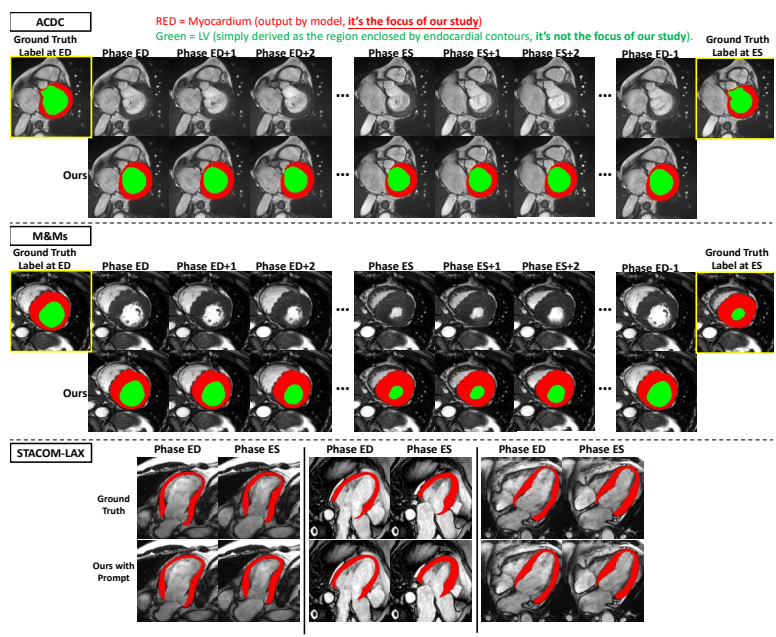

**Fig. 3.** The examples of our model's zero-shot performance. Ground truth labels at ED and ES for ACDC and M&Ms examples are shown in yellow box. Note that green represents left ventricle (LV) which is put here only for illustration. Our study focuses on the myocardium.

## 4       Conclusion

In this study, we propose cineCMR-SAM which employs SAM ViT blocks with pretrained SAM weights and temporal-spatial adaptation. This model delivers accurate and consistent segmentation of myocardium throughout all cardiac phases within a single cardiac cycle in cine CMR scans, outperforming SOTA methods both in data-specific models and in zero-shot generalization on unseen datasets. Incorporating text prompts which specifies the input view type, our model adeptly manages segmentation across both SAX and LAX views, laying groundwork for future adaptability to diverse inputs. More detailed text prompts such as "Basal SAX" or "three-chamber LAX" should be investigated in the future. We also plan to add regularization in training loss based on known cardiac behaviors such as smooth temporal motion and decreasing volume size during systole[9] in the future work. The validation study using multi-center in-house clinical CMR datasets and assessment of the clinical parameters derived from the segmentation results is currently ongoing.